# An empirical comparison of primary baffle and vanes for optical astronomical telescope

Taoran Li
Yingwei Chen

# An empirical comparison of primary baffle and vanes for optical astronomical telescope

Taoran Li, Yingwei Chen

Key Laboratory of Optical Astronomy, National Astronomical Observatories Chinese Academy of Sciences, 100012, Beijing, China;

## ABSTRACT

In optical astronomical telescopes, the primary baffle is a tube-like structure centering in the hole of the primary mirror and the vanes usually locate inside the baffle, improving the suppression of stray light. They are the most common methods of stray light control. To characterize the performance of primary baffle and vanes, an empirical comparison based on astronomical observations has been made with Xinglong 50cm telescope. Considering the convenience of switching, an independent vanes structure is designed, which can also improve the process of the primary mirror cooling and the air circulation. The comparison of two cases: (1) primary baffle plus vanes and (2) vanes alone involves in-dome and on-sky observations. Both the single star and the various off-axis angles of the stray light source observations are presented. The photometrical images are recorded by CCD to analyze the magnitude and the photometric error. The stray light uniformity of the image background derives from the reduction image which utilizes the MATLAB software to remove the stars. The in-dome experiments results reveal the effectiveness of primary baffle and the independent vanes structure. Meanwhile, the on-sky photometric data indicate there are little differences between them. The stray light uniformity has no difference when the angle between the star and the moon is greater than 20 degrees.

## 1. INTRODUCTION

Baffle and vanes are mechanical parts that block the propagation of unwanted light from source to the detector[1]. They are the most common methods of stray light control. The primary baffle is a tube-like structure, conical or cylindrical objects designed to block scattered lights, normally located above the hole of the primary mirror of an optical astronomical telescope. The vanes are usually added inside the primary baffle, used to block the lights scattered by the surface of the primary baffle and avoid the critical surface to be seen directly by focal plane.

The primary baffle will prevent large amount of the stray light. It could be seen on almost all the optical telescopes. While some telescopes such as SDSS[2] or LCOGT[3] only use the vanes without the primary baffle. Generally, the performance of the primary baffle is better than vanes but there are no empirical tests to characterize these two methods.

This paper describes the empirical comparison of primary baffle and vanes with Xinglong 50cm telescope, which aims to determine the facility selection of the stray light control. The campaign involves in-dome experiments and on-sky experiments, including (1) primary baffle plus vanes and (2) vanes alone.

## 2. THE INDEPENDENT VANES STRUCTURE

### 2.1 Xinglong 50cm telescope

The Xinglong 50cm telescope, located at Xinglong Observatory, is a classical Cassegrain telescope, with a primary mirror of diameter 50cm. As shown in figure 1, the Xinglong 50cm telescope has the open truss form and the fork mount. Only the stray light of small off-axis angles entering through the shutter of the hemispherical dome can illuminate the telescope. The instrument is PI 1300B, with the active pixels 1340 x 1300 and the pixel size 20 micron.

There are two advantages for making the comparison on this telescope:

1) It's easy to design the vanes due to the diameter of the original primary baffle is only 140mm;

2) Because of the height between the primary mirror and the floor is approximate 1.3 meters, the primary baffle and vanes are convenient to assemble or disassemble.

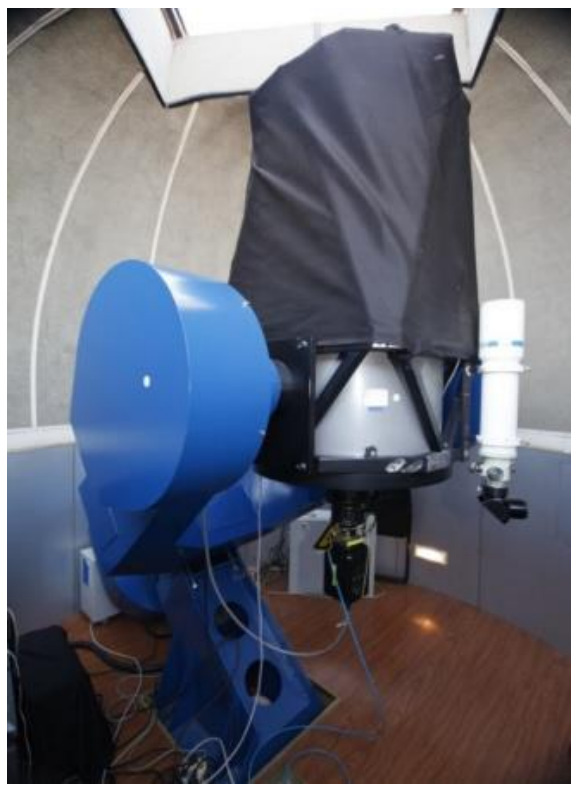

Figure 1 Xinglong 50cm telescope, with the fork mount in blue and the gray hemispherical dome. The open truss structure is covered by a black cloth to block the stray light and dust.

### 2.2 The independent vanes structure

The vanes are generally located inside inner surface of the primary baffle, screwing, sticking or integral forming. It’s hard to separate them during a short time when we need to compare their performance in a relatively steady atmosphere condition. Therefore, an independent vanes structure is designed, as shown in figure 2, which has the following advantages:

1) Light in weight;

2) Low cost;

3) Convenient to assemble or disassemble, using 3 screws to fix it;

4) Convenient to change between the case of primary baffle plus vanes and the case of vanes alone, the primary baffle can just be inserted on the vanes structure;

The independent vanes structure contains 7 vanes, 6 threaded rods and a pedestal. The location of each vane is fixed by 12 nuts. The nuts can be moved handily on the threaded rods which connect the 7 vanes. Each vane has been blackening to make it more absorptive. The rods are covered by black heat-shrink tubing because the threaded parts are hard to be blackening.

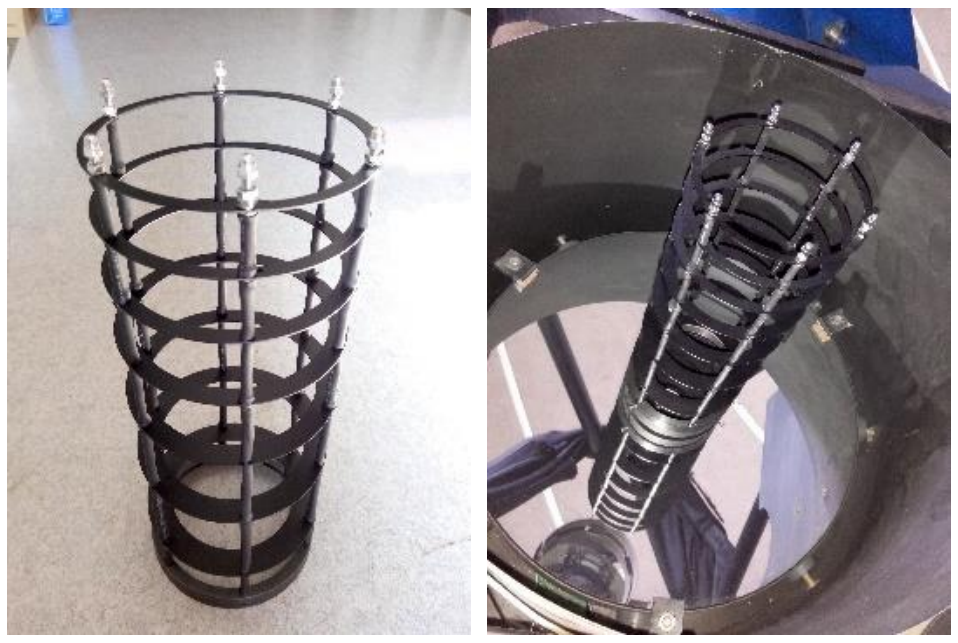

Figure 2 The independent vanes structure on Xinglong 50cm telescope. Left: 7 vanes connected by 6 threaded rods and fixed by 12 nuts; Right: The independent vanes structure on Xinglong 50cm telescope, using 3 screws to fix the pedestal and the central tube of primary mirror.

A study of this effect based on CFD (Computational Fluid Dynamics) method has shown that, when the airflow passes through the primary baffle, the “circular cylinder phenomenon” will occur behind the primary baffle, which will cause the turbulence above the mirror and increase the image FWHM (Full Width at Half Maximum). The vanes have advantages in respect of mirror cooling and air circulation[4].

# 3. IN-DOME EXPERIMENTS

The dark condition of the closed dome allows us to do the experiments during daylight and save the regular observation times. We take a light source on the wall with the 6 lamps of the dome as the stray light source, which simulate the strong illumination, point the telescope to zenith and take off the black cloth, as shown in figure 3.

In-dome experiments consist of 3 cases: (1) primary baffle plus vanes; (2) only vanes; (3) no primary baffle and vanes. During the experiments, the telescope holds the pointing without tracking. The exposure time is 1 second. The raw fit images are imported to MATLAB and analyzed with it.

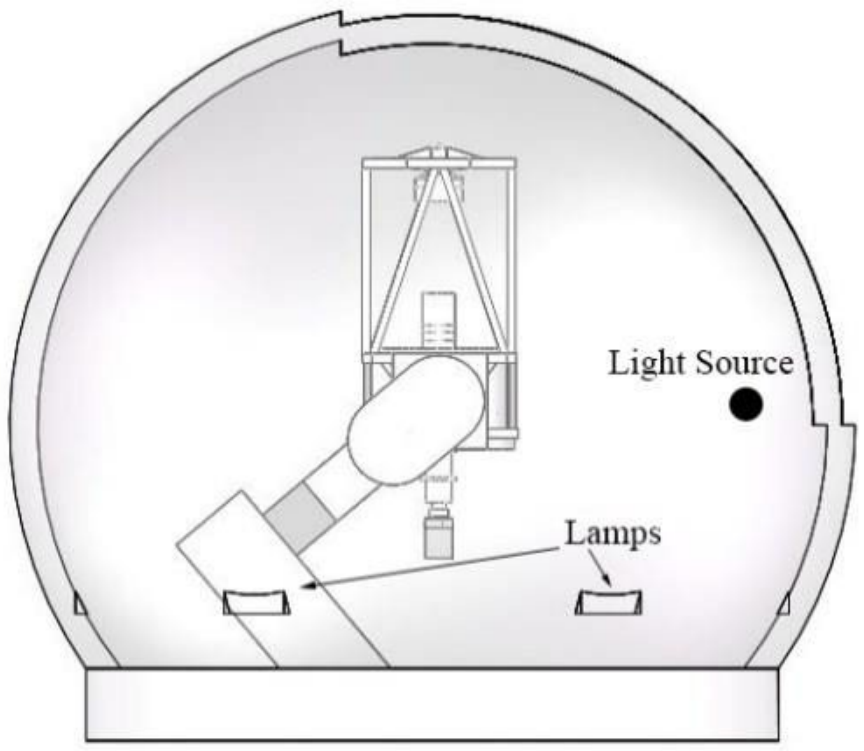

Figure 3 The light source and lamps inside the dome. The light source is a stable source with a power about 1 watt. The height between the light source and floor is about 1.5 meters. The 6 lamps locate on the wall of the dome, symmetrically. The telescope is pointed to zenith and the dome shutter is closed.

## 3.1 Contour map

Figure 4 is the contour maps on CCD in the 3 cases. The different colors represent different ADU (Analogue-to-Digital Unit) value. In MATLAB, the ADUs are converted to colors and the range of colors is calculated by the maximum and minimum of ADUs. The contour map allows us to check the ADU distribution clearly. From the CCD center to the edge, the ADUs reduce gradually in all cases. The case 1 (primary baffle plus vanes) has the lowest ADU value and standard deviation, as shown in Table 1, that means the stray light suppression in case 1 has the best performance.

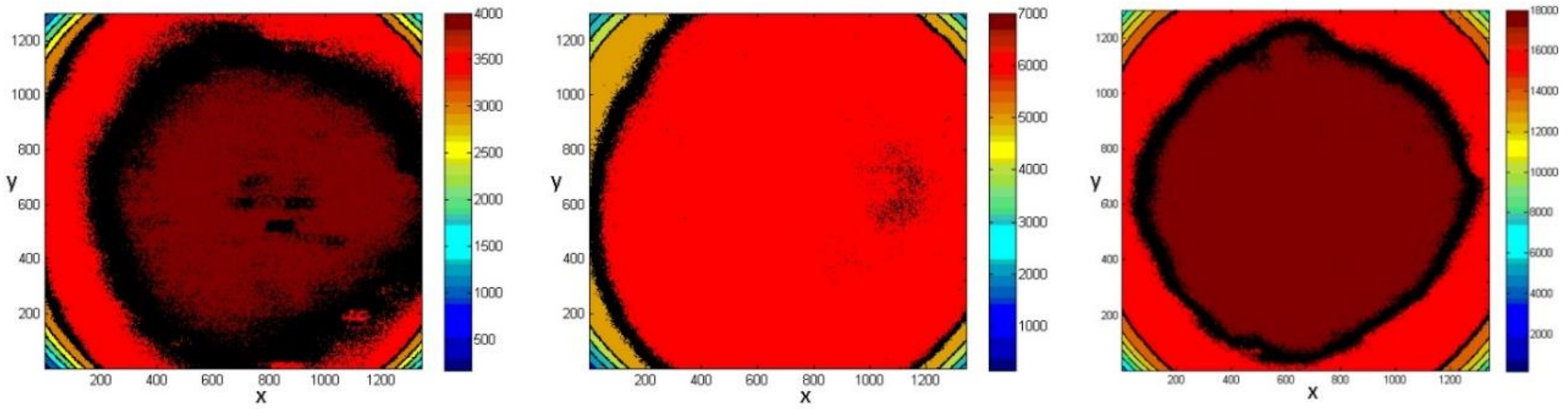

Figure 4 Contour maps on CCD. Left: Primary baffle plus vanes; Middle: Only vanes; Right: No primary baffle and vanes. The black circles represent the contour line which is used to isolate the area of each color (value).

Table 1 Maximum, median and standard deviation of ADU

| ADU | Maximum | Median | Standard deviation |
|---|---|---|---|
| no primary baffle and vanes | 19429 | 18187 | 1269 |
| only vanes | 7102 | 6549 | 494 |
| primary baffle plus vanes | 4360 | 4024 | 314 |

## 3.2 Histogram

A histogram is a graphical representation of the distribution of ADUs. If the stray light distribution of the image is uniform, the ADUs shall distribute in one section. Therefore, the uniformity of stray light can be examined roughly from the histogram. Due to the fit imported to MATLAB is a 2D matrix, it needs to be converted to 1D matrix, which is 1*1742000 (1340 x 1300=174200).

Figure 5 is the histogram of 3 cases, with an interval of the histogram of 100. The x-axis is ADU value and y-axis is the frequency. The frequency of each interval is recorded in Table 2, which contains the ADUs range, frequency and percentage.

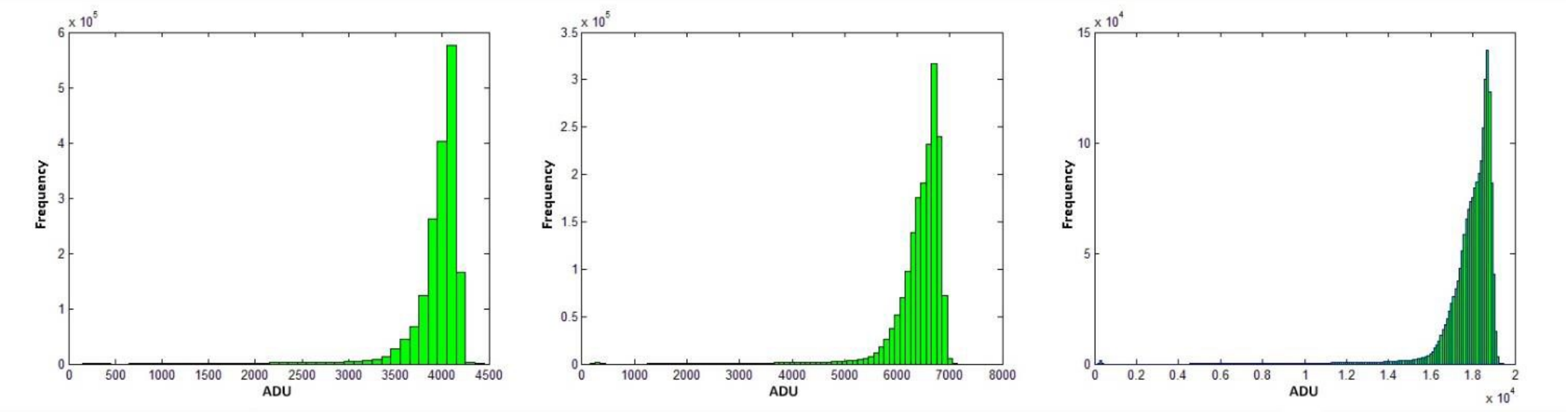

Figure 5 Histogram of ADUs on CCD. The x-axis is ADU value and y-axis is the frequency. The interval of the ADU is 100. Left: Primary baffle plus vanes; Middle: Only vanes; Right: No primary baffle and vanes.

Table 2 The ADU range which has the most frequency and percentage

| | ADU range | Frequency | Percentage |
|---|---|---|---|
| Primary baffle plus vanes | 4000-4100 | 576913 | 33.12% |
| Only vanes | 6700-6800 | 317133 | 18.21% |
| No primary baffle and vanes | 18700-18800 | 141736 | 8.14% |

# 4. ON-SKY EXPERIMENTS

The on-sky experiment is the best method to test the stray light performance of one telescope, which has absolutely the same environment as the astronomical observations. The stray light observations utilize the photometrical data to analyze the sky value, magnitude and stray light uniformity.

## 4.1 Single target

The target of experiment is Landolt 111_1965, the magnitude in V band is 11.4. The exposure time is 45s. The angle between the target and the moon is about 26 degrees. The moon phase is 35% and the atmosphere conditions are stable. Table 3 and figure 6 show the angle, azimuth and altitude of the target and the moon.

The experiments contain the cases of (1) primary baffle plus vanes and (2) only vanes. Because of the short exposure time and alternate time, it can be treated that the locations of the target and moon remain the same, that is, the off-axis angle of stray light source is not changed during the experiments.

Table 3 The azimuth and altitude of the target and the moon

| Landolt 111_1965 | $RA: 19h37m42s \ DEC: +00^\circ 26'30''$ |
|---|---|
| Time (UT) | 10:52:56 |
| Target Az and Alt | $Az: 207^\circ 47'20'' \quad Alt: 42^\circ 24'18''$ |
| Moon Az and Alt | $Az: 217^\circ 39'24'' \quad Alt: 18^\circ 24'08''$ |

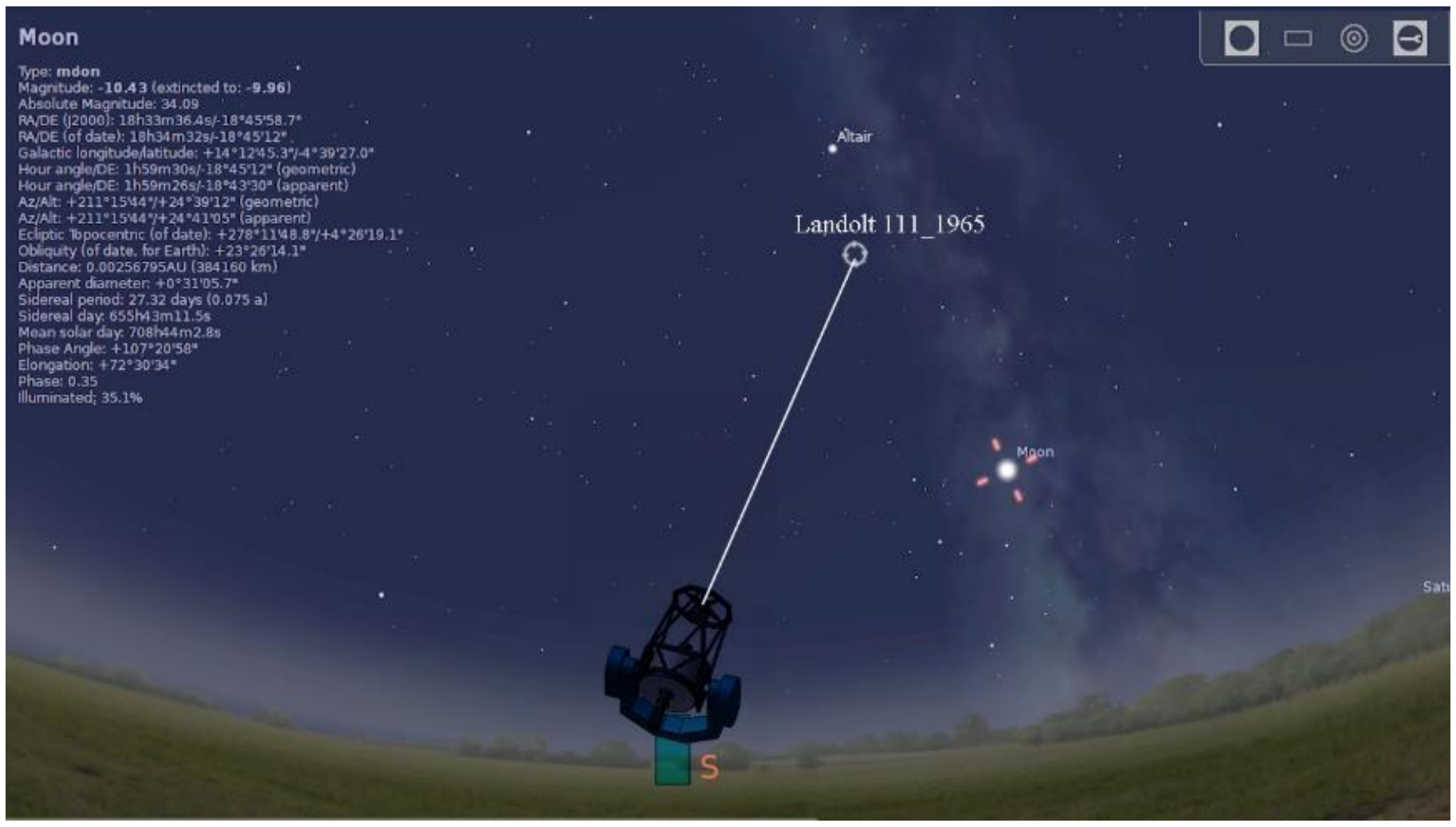

Figure 6 Sketch map of the angle between the telescope pointing direction and the moon.

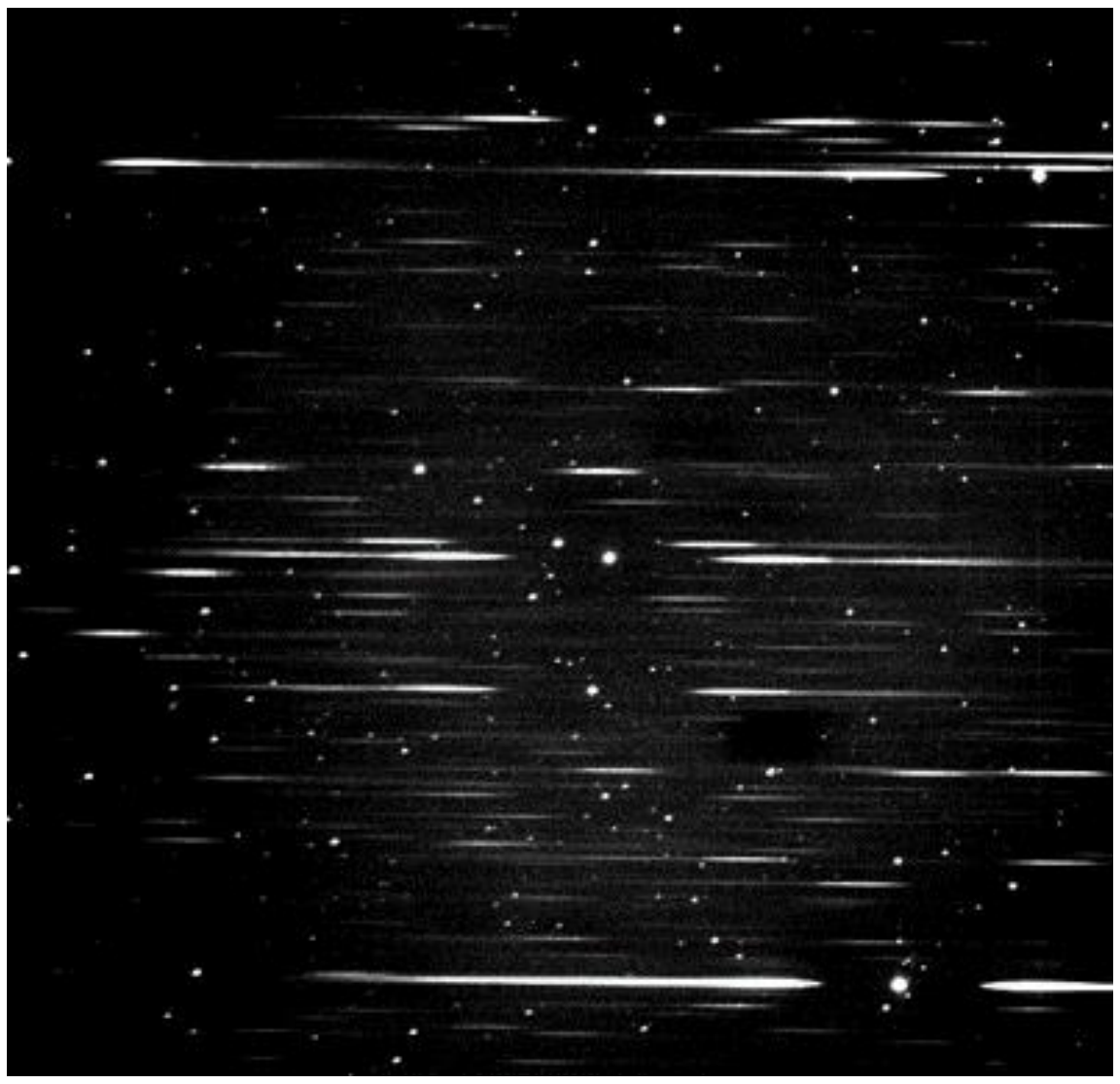

Figure 7 The raw image of one star field. There are many first order spectra on CCD image.

Due to the grism in front of the CCD, there are first order spectra on CCD image, as shown in figure 7. We use the zero order spectra to do the photometry. The IRAF (Image Reduction and Analysis Facility) software is used to do the aperture photometry on both 111_1965 and 111_1969 to compare the sky background value and the magnitude, as shown in Table 4. The photometrical results from IRAF illustrate that the sky background values for case 2 (only vanes) are a little higher than case 1 (primary baffle plus vanes), but the magnitude differences are so small and can be negligible.

Table 4 Photometrical results from IRAF

| **Unit: counts** | **Sky** | **Mag** |
|---|---|---|
| Primary baffle plus vanes | 2053.56 | 11.33 |
| Only vanes | 2054.36 | 11.26 |
| Primary baffle plus vanes | 2061.51 | 9.79 |
| Only vanes | 2074.50 | 9.78 |

To archive the stray light uniformity (the distribution of background stray light on CCD), we need the contour map and the histogram. However, the intensity of zero and first order spectra is much larger than the background, the contour map will be filled up by one color that represent the intensity of the star. It's impossible to examine the uniformity (Figure 8).

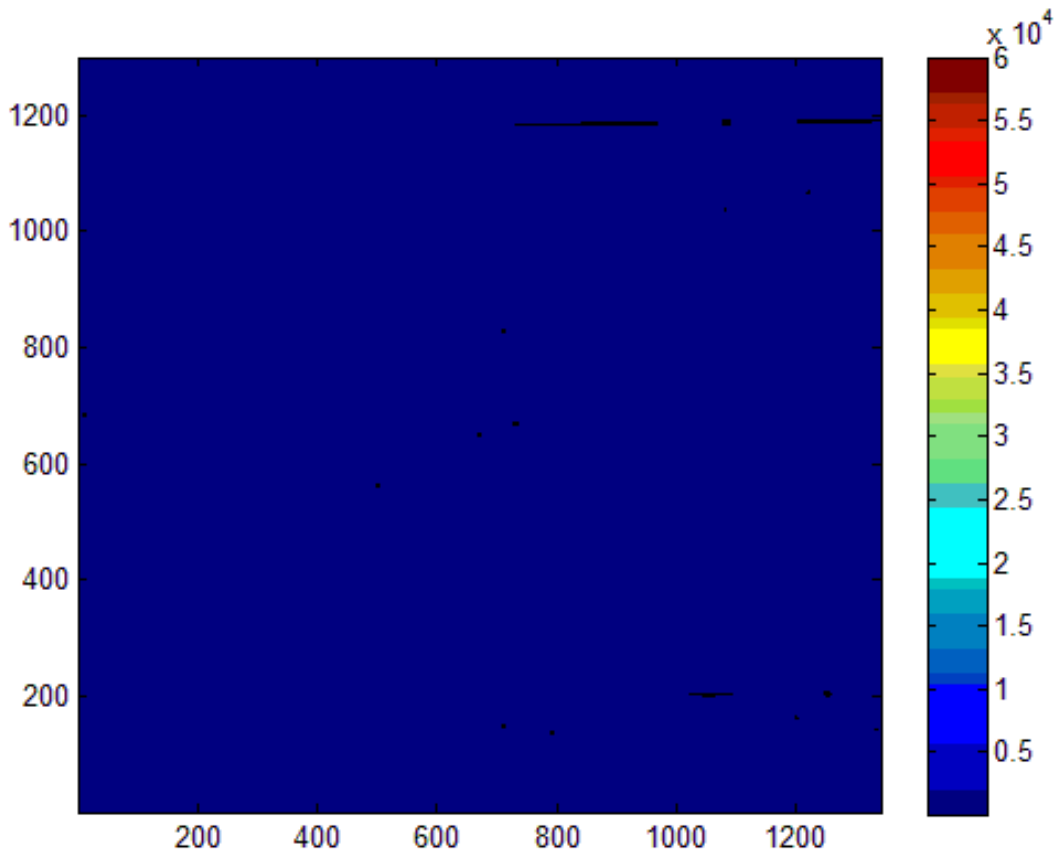

Figure 8 Contour map on CCD, with the zero and first order spectra

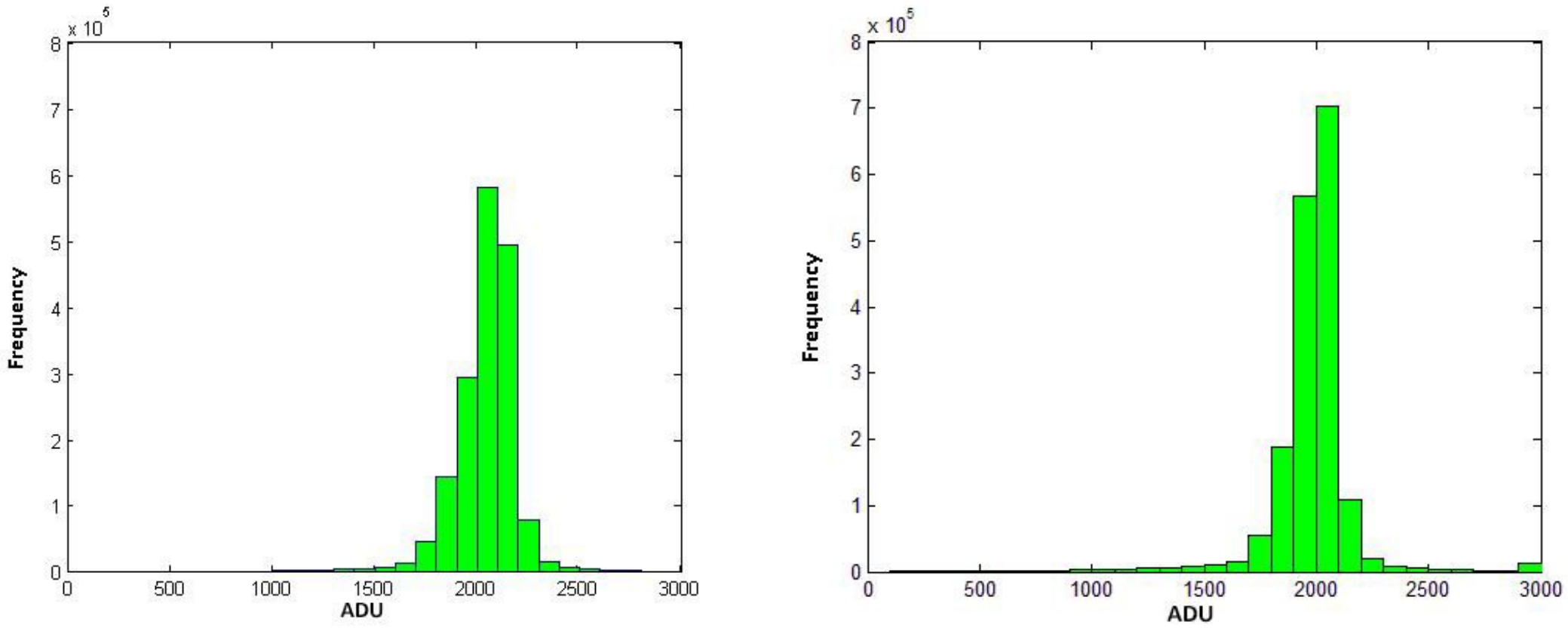

Figure 9 ADU histogram. Left: Primary baffle plus vanes. Right: Only vanes

Here, we define a simple method to delete the stars (zero and first order spectra) from the images with MATLAB. The first step is to draw a histogram, the same as the method used in section 3.2. As the histogram with an interval of the histogram of 100 shown in figure 9, most ADUs locate in the range of 1000 to 2500. The ranges which have the most frequency and the percentage are recorded in Table 5. Both the histogram and table show that, the range of the biggest frequency is 2000-2100 and the case 1 (primary baffle plus vanes) has a larger frequency, it is more concentrated. In this case, the stray light uniformity is better.

Table 5 The ADU range which has the most frequency and percentage

| | ADU range | Frequency | Percentage |
|---|---|---|---|
| Primary baffle plus vanes | 2000-2100 | 702981 | 40.35% |
| Only vanes | 2000-2100 | 580736 | 33.34% |

The second step is to regard the points which have the low frequency and the bigger ADU value as the zero and first spectra of the star. That is, for the two cases, regard the ADU larger than 2500 as the spectra and less than 2500 as the background stray light.

The final step is to reduce the range of the color bar because MATLAB is unable to distinguish the color if the range of the color bar is too large (figure 8). As shown in figure 9, the ADUs less than 1000 are such a small quantity that it doesn't affect the background distribution. Therefore, let the ADUs larger than 2500 equal to 1000, the color in MATLAB can be distinguished easily. Only the stray light (background value) are remained in the contour map (figure 10). We can get the conclusion from the contour maps quickly and qualitatively, in the instance of this moon phase and the telescope pointing, the stray light uniformity of the case 1 is better than case 2, but not much difference.

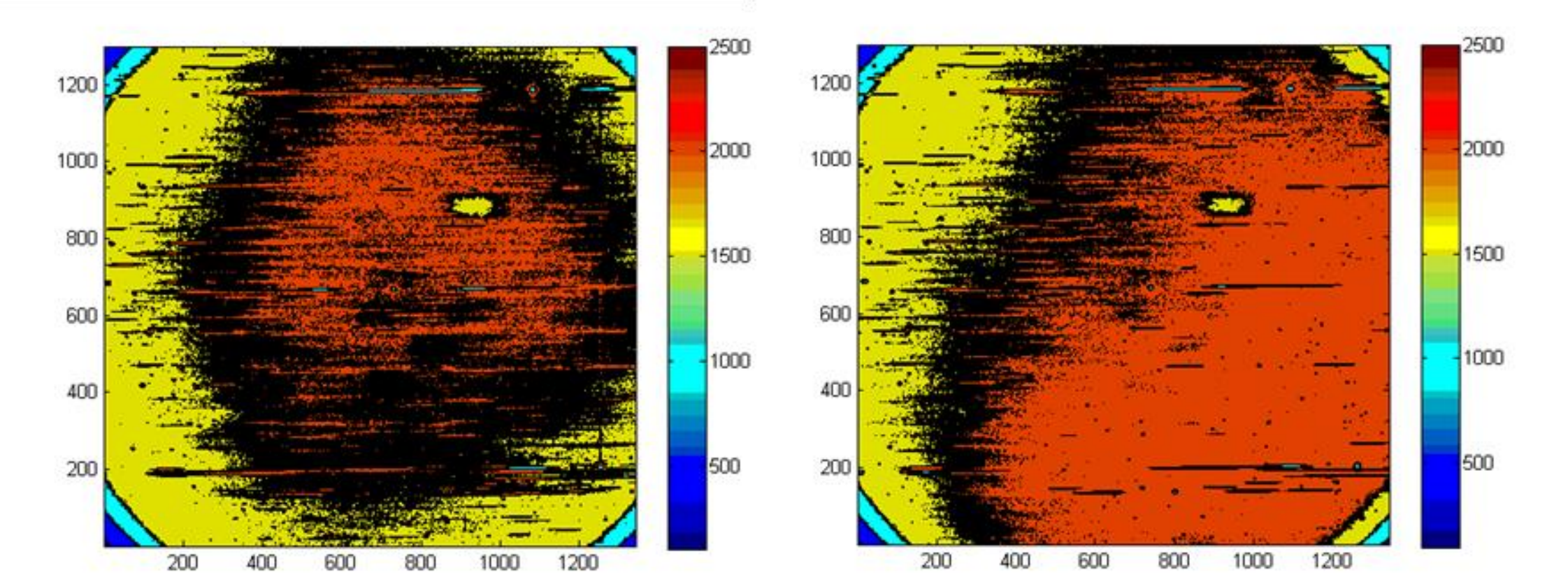

Figure 10 Contour map after deleting the star. Left: Primary baffle plus vanes. Right: Only vanes

### 4.2 Observations with a moving stray light source

A moving stray light source will cause the varied stray light performance of one telescope. In this situation, the performance of the primary baffle and vanes remains to be tested. To improve the efficiency of the experiments and focus on comparison of primary baffle and vanes, the moon is regarded as the stray light source and the telescope is pointed to 5, 10, 20, 30, 40, 50, 60 degrees away from the moon, which makes a "moving moon" (during the experiments, the exposure time and the slewing time is much short that the moon can be treated as a fixed source), as shown in figure 11. The moon phase is 26% and the atmosphere conditions are stable. The experiments still contain two cases: (1) primary baffle plus vanes, (2) only vanes.

The photometric data-magnitude, sky background value and standard deviation for one target in each angle are acquired using IRAF CCD reduction, with bias and flat corrections. Figure 12 is the sky background value and the standard deviation curves. The angle increasing, the value decreasing. When the angle is less than 10 degrees, there will be a quite big difference in both two images. In this range, the primary baffle is more efficient. After 20 degrees, the differences can be ignored.

Figure 13 is the line chart and the D-value of the magnitude. The red and blue lines are the magnitudes of the case 1 and 2, respectively. The magnitudes of two cases are nearly the same. The max D-value is 0.085 at 20 degrees, others are less than 0.04.

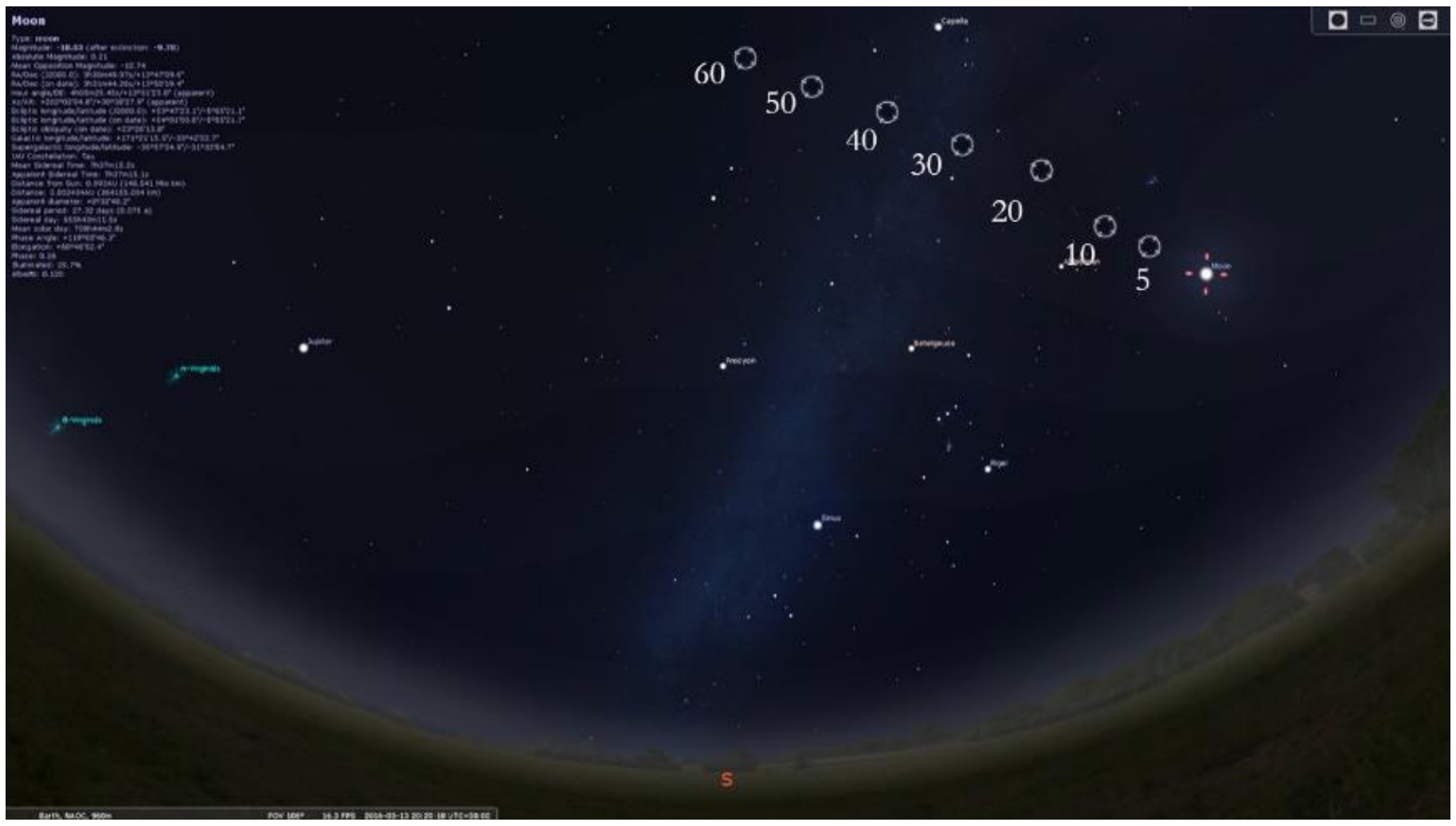

Figure 11 The angles between the telescope and the moon.

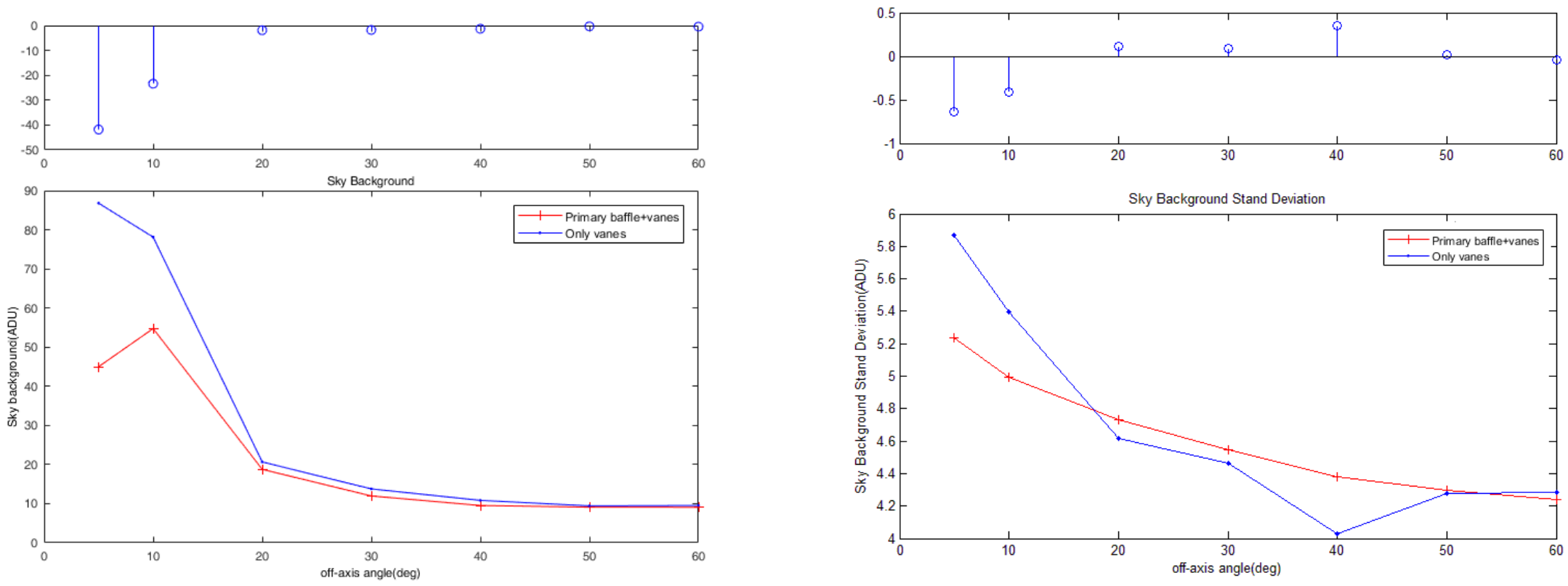

Figure 12 Sky background value (left) and standard deviation (right). The angle increasing, the value decreasing. After 20 degrees, the differences of sky background and standard deviation can be ignored.

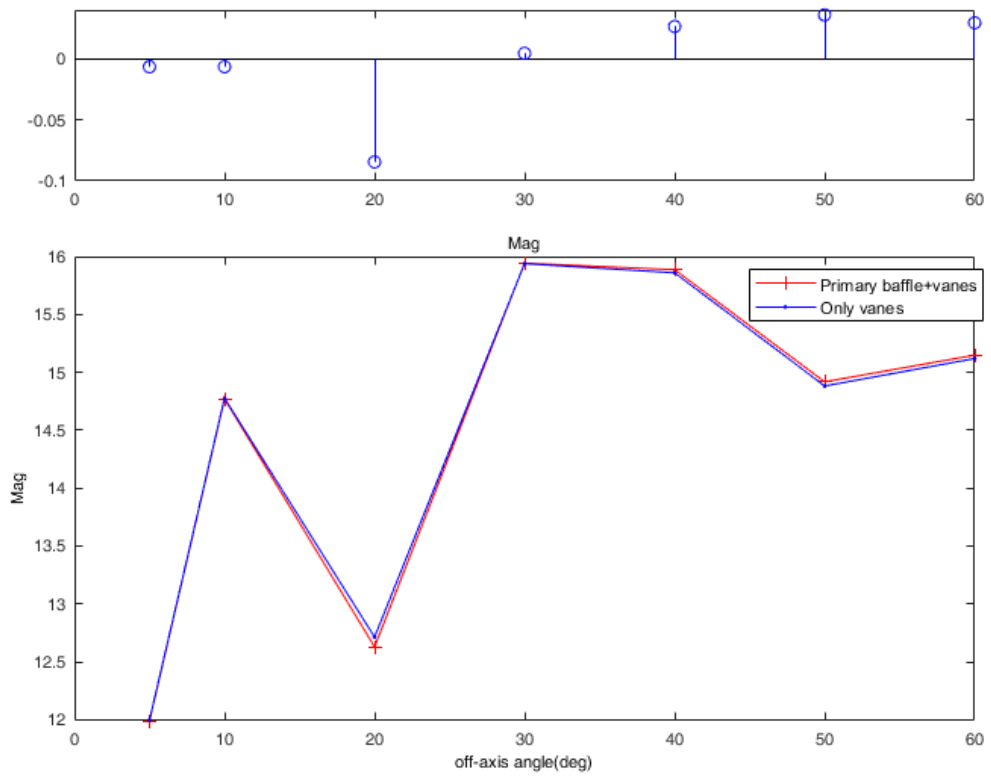

Figure 13 The comparison of photometric magnitude, x axis is the distance to the moon, y is the magnitude. The red and blue lines are the magnitudes of the primary baffle plus vanes and only vanes, respectively. Only at 20 degrees, the difference is almost 0.1, others are less than 0.04.

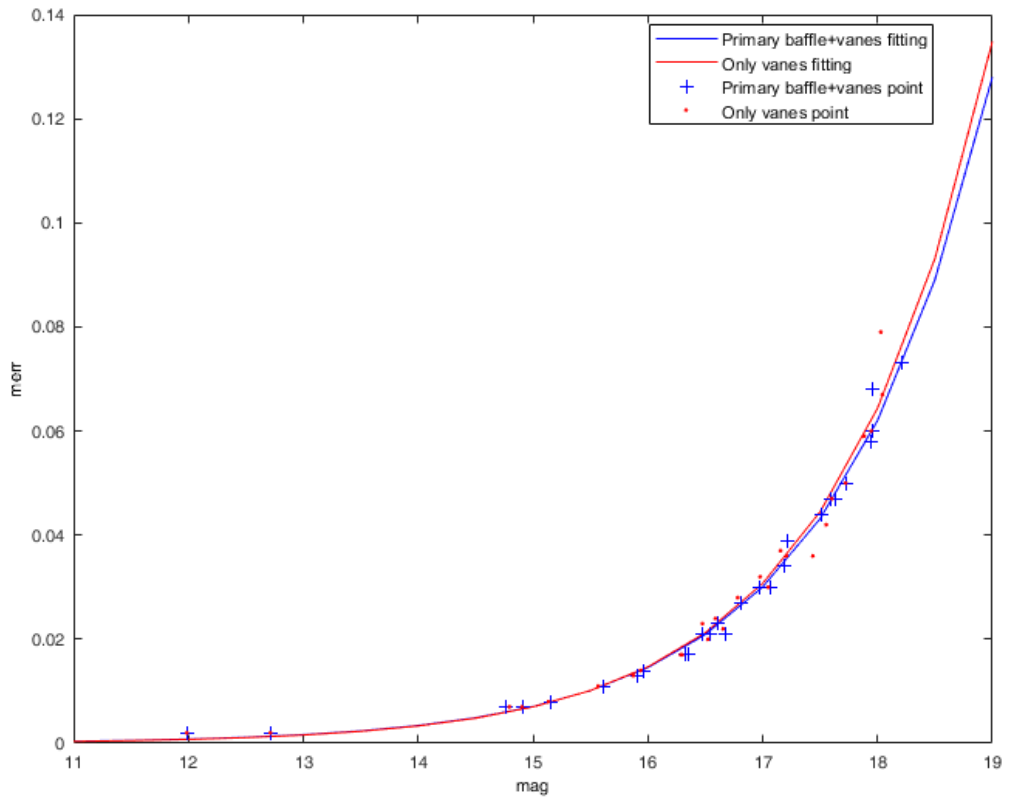

Figure 14 The magnitude-errors curves of primary baffle plus vanes (blue line and points) and only vanes (red line and points).

The magnitude errors (expressed by $\sigma_m$ ) is an important factor of the telescope, indicates the photometric accuracy, which is related to the Signal to Noise Ratio (SNR)[5],

$$\sigma_m = 2.5 \lg(1 + 1/SNR) \tag{1}$$

The magnitude-error diagram is obtained from the photometric data of 3~5 stars on the images of each angles, as shown in figure 14. The blue and red lines represent the fitting curves for case 1 and 2, respectively. The cross and dot marks show the data points. The errors are almost no difference only if the star is faint. For instant, the difference of the photometric error of case 1 and 2 is 0.22% if the instrumental magnitude is 17.95. As a star with 16 instrumental magnitude, the difference is only 0.02%.

Table 6 Instrumental magnitude and photometric error of primary baffle plus vanes, only vanes

| Instrumental magnitude | Error (case 1) | Error (case 2) | Difference |
|---|---|---|---|
| 17.95 | 5.98% | 6.20% | 0.22% |
| 16.00 | 1.45% | 1.47% | 0.02% |

The method for background test is the same as section 4.1-delete the star. The CCD binning[6] allows charges from adjacent pixels to be combined, which can offer benefits in computer processing speed and this will not affect the stray light distribution. The 2 x 2 binning is used twice in this work. The resolution of the image after binning is 335 x 325 (The original is 1340 x 1300). Figure 15 is the binning contour map of different angles with the zero and first spectra deleting. The first and third rows are the case-primary baffle plus vanes, the second and fourth rows are the case-only vanes. The angles from moon are 5,10,20,30,40,50,60 degrees. The different colors represent different intensities of the background stray light. At 5 and 10 degrees, several divisions appear clearly on the images of case 2. The intensity of the background is getting smaller from inside to outside. When the angles are less than 20, the differences are obvious, which means the primary baffle is more efficient in stray light distribution improvement. After 30 degrees, the case 1 and 2 are almost the same.

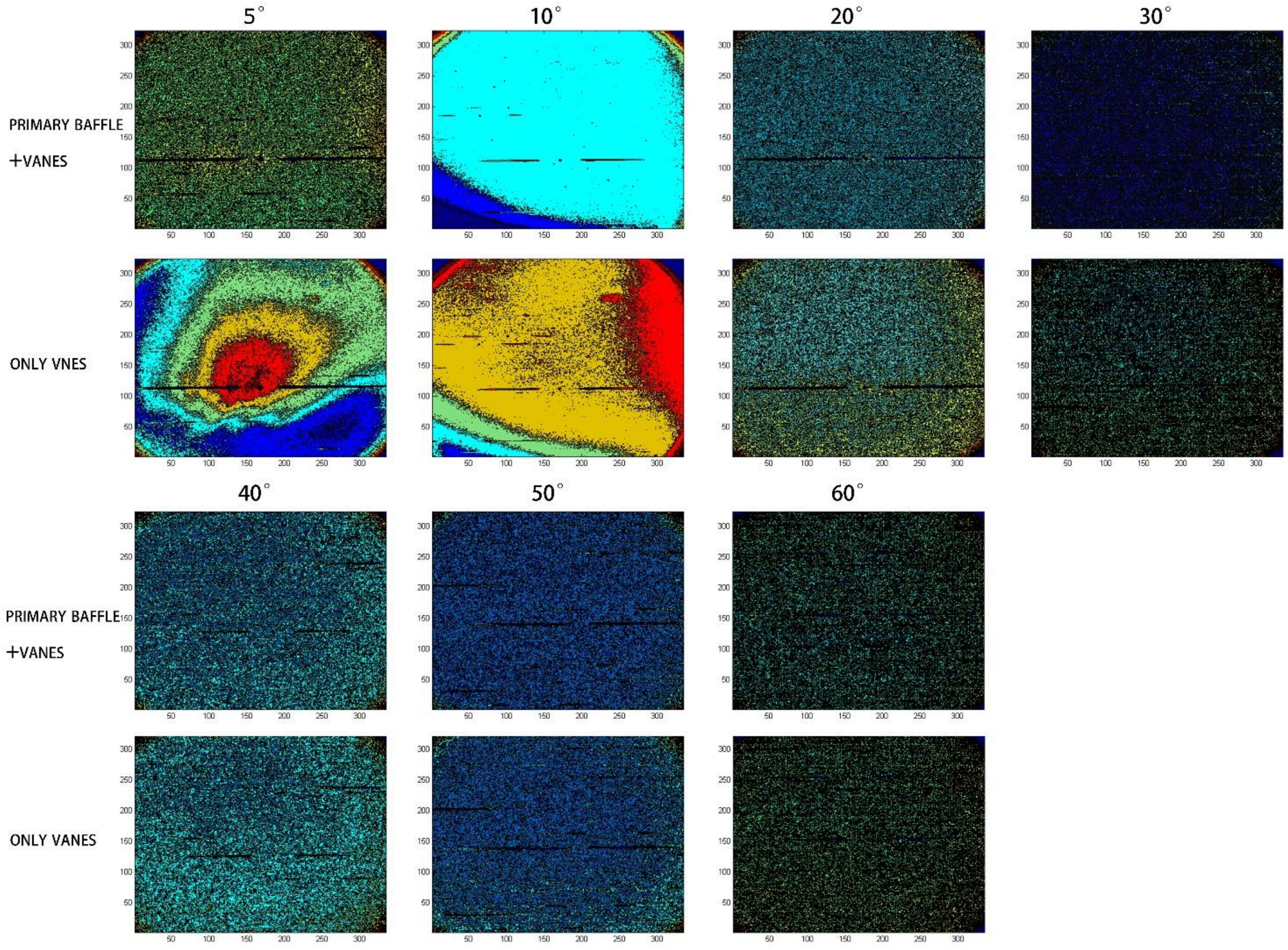

Figure 15 Contour map of different angles with the zero and first spectra deleting. The first and third rows are the case-primary baffle plus vanes, the second and fourth rows are the case-only vanes. From the up to the bottom, the angles from moon are 5,10,20,30,40,50,60 degrees, respectively.

## 5. CONCLUSIONS

The empirical comparison of primary baffle and vanes is based on stray light observation, which involves an analysis of the magnitude, photometric error and the stray light uniformity. Considering the convenience of switching between primary baffle and vanes, we have designed an independent vanes structure and focused on a 50cm telescope equipped with (1) primary baffle plus vanes and then (2) vanes alone, examined the performance of stray light and background uniformity.

In strong illumination (in-dome experiments), both the primary baffle and independent vanes structure are effective at suppressing stray light. The case “primary baffle plus vanes” has the best suppression performance. During the on-sky experiments, we observe big differences only when the moon is within 30 degrees of the observed star. The difference of the photometric error between case 1 and 2 is only 0.02% when the instrumental magnitude is 16.

Given these results, and the needs of typical astronomical observation, the independent vanes structure can be used without the primary baffle to improve the mirror cooling and the air circulation when telescope points to a bright star, or the telescope is 30 degrees away from the moon.

## ACKNOWLEDGEMENTS

The authors gratefully acknowledge the supports of the team of Xinglong Observatory: Xiaojun Jiang, Jianfeng Wang, Jiupeng Guo, Zhigang Hou and Liguo Fang. Contribution from Chris Benn of the Isaac Newton Group of Telescopes also is gratefully acknowledged.